\documentclass[a4paper,11pt]{article}
\usepackage{authblk}
\usepackage{graphicx}
\title{Universality of the Majorana Double Charge Exchange}

\author[1,2]{C. Garofalo}
\author[3]{H.Lenske}
\author[1,2]{F. Cappuzzello}
\author[2]{M. Cavallaro}
\affil[1]{Dipartimento di Fisica e Astronomia 'Ettore Majorana', Università di Catania, Catania, Italy}
\affil[2]{Istituto Nazionale di Fisica Nucleare, Laboratori Nazionali del Sud, Catania, Italy}
\affil[3]{Institut fur Theoretische Physik, Justus–Liebig–Universitat Giessen, D-35392 Giessen, Germany}

\date{}
\begin{document}

\maketitle

\begin{abstract}
The Majorana Double Charge Exchange nuclear reaction mechanism represents a powerful and novel scenario to probe the dynamics of the neutrinoless double beta decay. Recent studies presented in this manuscript have highlighted a key feature of this mechanism, namely its independence with respect to the nuclei involved in the reaction.
\end{abstract}


\section{Introduction}
Double Charge Exchange (DCE) reactions are second-order nuclear processes in which the projectile and target nuclei exchange two units of charge, two units of isospin, as well as energy and momentum, while the mass number remains unchanged:
\begin{equation}
    ^a_zx\,\, +\,\, ^A_ZX\longrightarrow^a_{z\pm 2}y \,\, +\,\,^A_{Z\mp 2}Y
\end{equation}

The DCE process can proceed through the sequential transfer of two protons and two neutrons between projectile and target. This mechanism, known as Transfer Double Charge Exchange (TDCE) \cite{ferreira2022multinucleon, ferreira2025analysis}, is driven by the mean field. Moreover, the reaction can occur through the exchange of charged mesons between either uncorrelated or correlated nucleons, leading to the Double Single Charge Exchange (DSCE) \cite{bellone2020two, bellone2025heavy} and Majorana Double Charge Exchange (MDCE) \cite{lenske2024theory} mechanisms, respectively. The former corresponds to a two-step DCE process driven by the isovector nucleon–nucleon interaction $T_{NN}$, which acts twice as a one-body operator in both the target and projectile nuclei. The latter is a one-step DCE process mediated by the isovector pion–nucleon interaction $T_{\pi N}$, arising from off-shell pion–nucleon scattering. 

The MDCE mechanism is of great interest, as it presents a close connection with neutrinoless double beta ($0\nu\beta\beta$) decay \cite{majorana1937teoria, furry1939transition}. In both processes, the same initial and final many-body nuclear states are involved, while the corresponding transition operators present significant mathematical similarities. Indeed, both operators include Fermi, Gamow–Teller, and rank-two tensor components. In this context, DCE reactions provide a means to probe and put constrains  on the Nuclear Matrix Elements (NMEs) of $0\nu\beta\beta$ decay, currently embedded in a complex puzzle \cite{agostini2023toward}. This constitutes the main goal of the NUMEN (NUclear Matrix Element for Neutrinoless double beta decay) project \cite{cappuzzello2023shedding, cappuzzello2018numen}, proposed in 2015 to investigate the fascinating and challenging heavy-ion–induced DCE reactions.


\section{Majorana Double Charge Exchange mechanism}
\begin{figure}[htb]
\centerline{%
\includegraphics[width=5 cm]{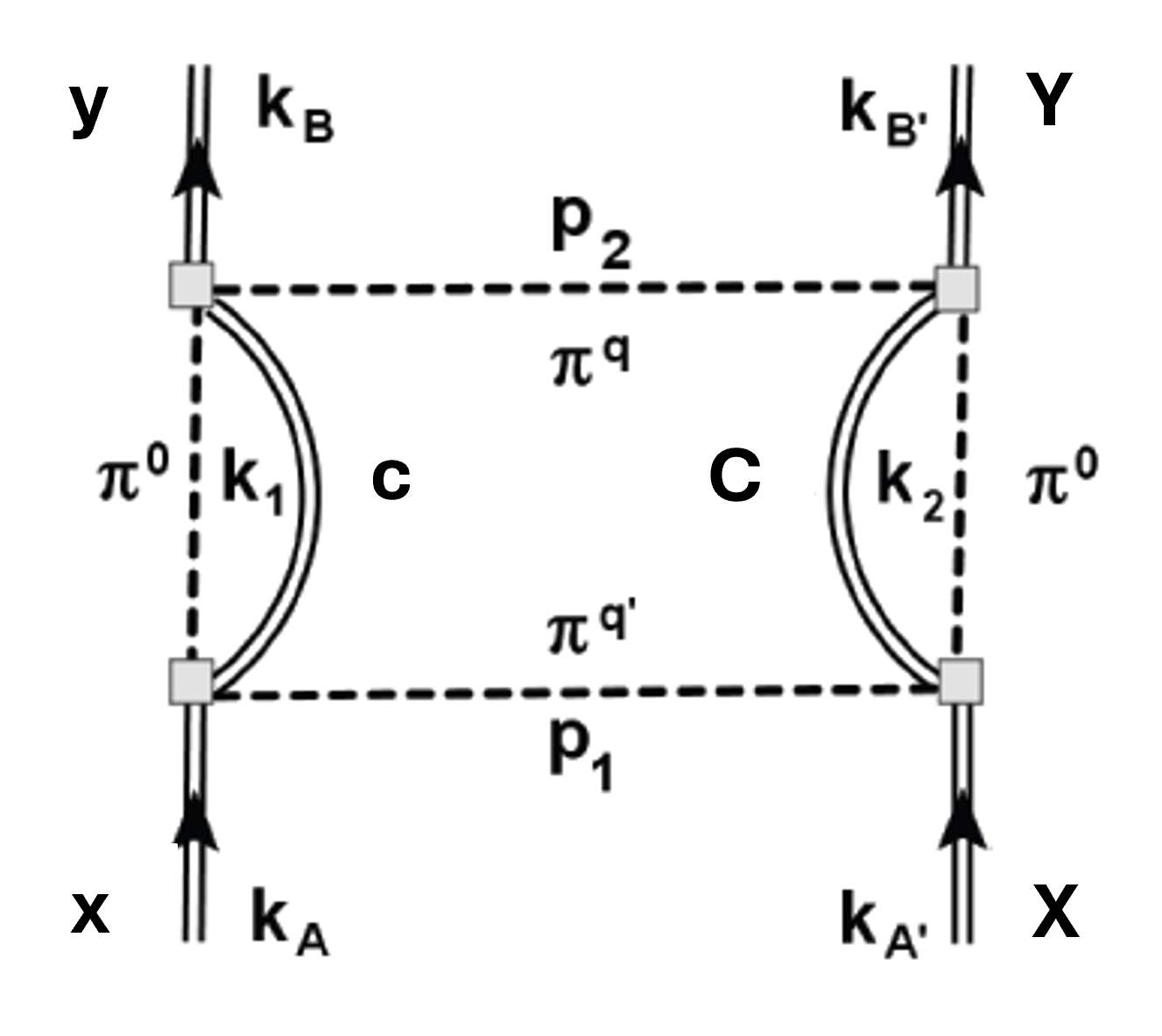}}
\caption{Scheme of the MDCE for the $X(x, y)Y$ reaction.}
\label{Fig:MDCE}
\end{figure}

The MDCE mechanism is described by the box diagram shown in Fig. \ref{Fig:MDCE}. In the incident channel, the target $x$ and the projectile $X$ exchange two charged pions $\pi^{\pm}$. The pion-nucleon scattering induces a Single Charge Exchange (SCE) - type transition in both nuclei, accompanied by the emission of a neutral pion $\pi^0$. The $c$ and $C$ nuclei together with the $\pi^0$ propagate through the intermediate channel until a $\pi^0 \rightarrow \pi^\pm$ conversion occurs, inducing a second SCE - type transition in the exit channel. 

\section{Pion potentials}
Using the pion rest mass $m_{\pi}\sim 139\,MeV$ as a natural separation scale, the pion potential entering the intrinsic nuclear DCE vertices \cite{lenske2024theory} can be evaluated in closure form:  

\begin{equation}
    U_{\pi}(x)= - \int \frac{d^3 k}{(2\pi)^3} T_{\pi N}(\vec{p}_2,\vec{k})\frac{e^{i\vec{k}\cdot\vec{x}}}{m_{\pi}^2+k^2} T_{\pi N}(\vec{p}_1,\vec{k}) \,\,\, ,
    \label{eq:U}
\end{equation}
where $x$ is the distance between the two nucleons participating in the DCE transition, $k$ is the invariant relative pion-nucleon momentum. In the region relevant for the MDCE process the (off-shell) isovector pion-nucleon T-matrix $T_{\pi N}$ is adequately described by the operator structure: 

\begin{equation}
    T_{\pi N} = \left[T_0(s_{\pi N}) + \frac{1}{m_{\pi}^2}{\left( T_1(s_{\pi N})\vec{p}\cdot\vec{p}' + iT_2 (s_{\pi N}) \vec{\sigma} \cdot(\vec{p}\times p') \right)}\right] \vec{T}_{\pi} \cdot\vec{\tau}_{N}
    \label{form:T-matrix}
\end{equation}
%
where $T_0$, $T_1$ and $T_2$ are the form factors describing the S- and P-wave interactions arising from the pion-nucleon scattering \cite{moorhouse1969pion, johnson1993pion}. The pion and nucleon isospin operators are denoted by $\vec{T}_{\pi}$ and $\vec{\tau}_{N}$, respectively. 

\begin{figure}[htb]
\centerline{%
\includegraphics[width=14 cm]{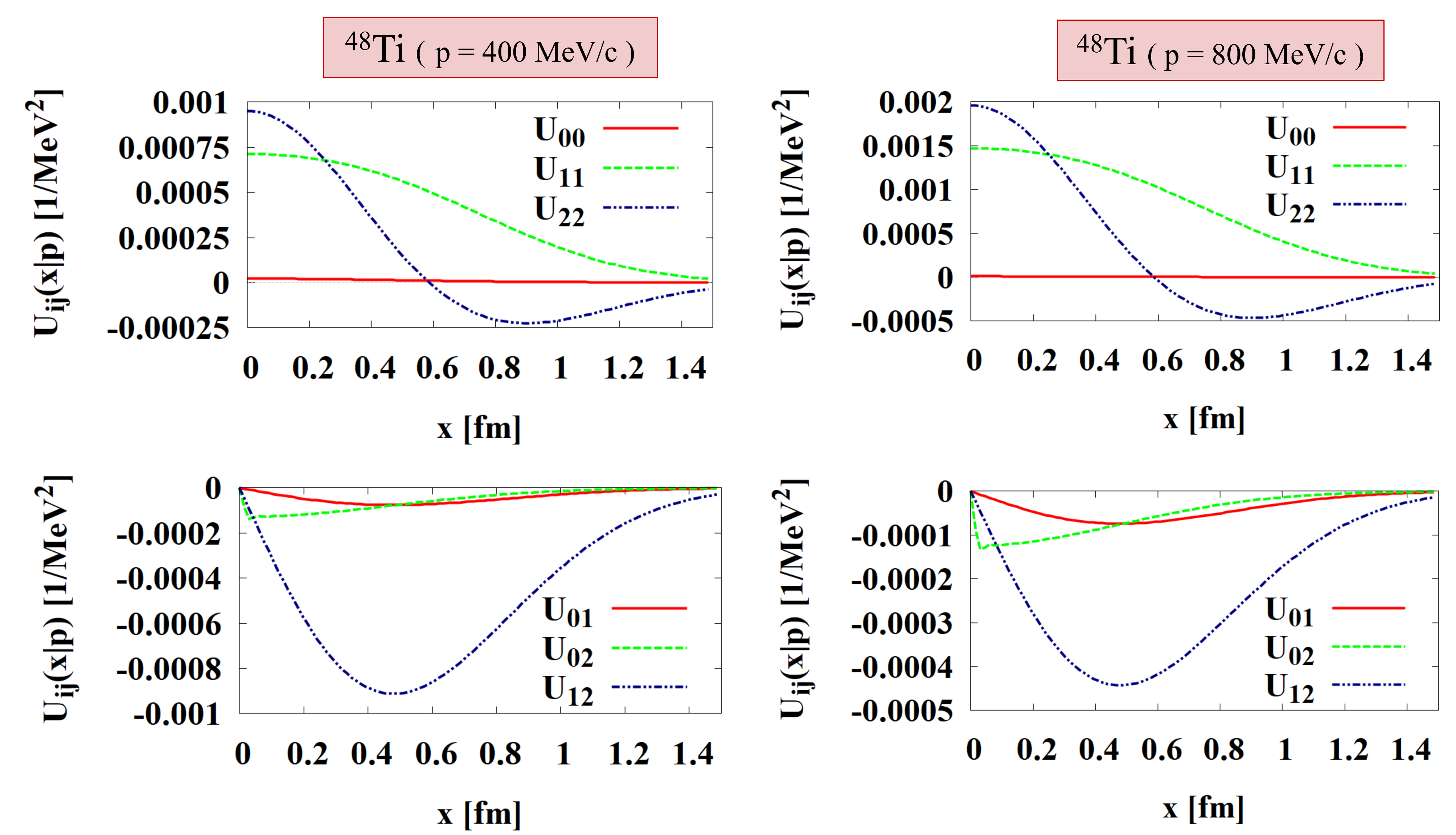}}
\caption{The different component of the MDCE pion potentials for collinear $\vec{p}_1 \parallel \vec{p}_2$ momenta at $|\vec{p}_1| = |\vec{p}_2| = 400$ MeV/c (left panel) and $|\vec{p}_1| = |\vec{p}_2| = 800$ MeV/c (right panel).}
\label{Fig:pion_potentials}
\end{figure}

The pion potential consists of nine terms, as it arises from the double product of $T_{\pi N}$ defined in Eq. \ref{form:T-matrix}, which itself is a superposition of three terms. Under collinear kinematics ($\vec{p}_1 \parallel \vec{p}_2$, $|\vec{p}_1| = |\vec{p}_2|$), only six independent pion potential components $U_{ij}$ are obtained. Typical results for $U_{ij}$ are shown in Fig. \ref{Fig:pion_potentials}. The pion potentials for the $^{48}$Ti nucleus coming from the collision $^{18}$O $+ \,^{48}$Ti at $270$ MeV are plotted as a function of the distance between the two nucleons involved in the MDCE reaction, exchanging a charged pion with momentum $p = 400$ MeV/c (left panel) and $p = 800$ MeV/c (right panel). The size of the nucleus affects the integration range in Eq. (\ref{eq:U}). As it can be seen in Fig. \ref{Fig:pion_potentials}, the P-wave potentials - namely, the diagonal components $U_{11}$ and $U_{22}$ - dominate over the S-wave ones, which remain small and moderate in magnitude compared to the P-wave parts at both the explored momenta. Therefore, it can be concluded that the MDCE process is predominantly governed by the P-wave component.

\section{Universality of the MDCE mechanism}
\begin{figure}[hb!]
\centerline{%
\includegraphics[width=14 cm]{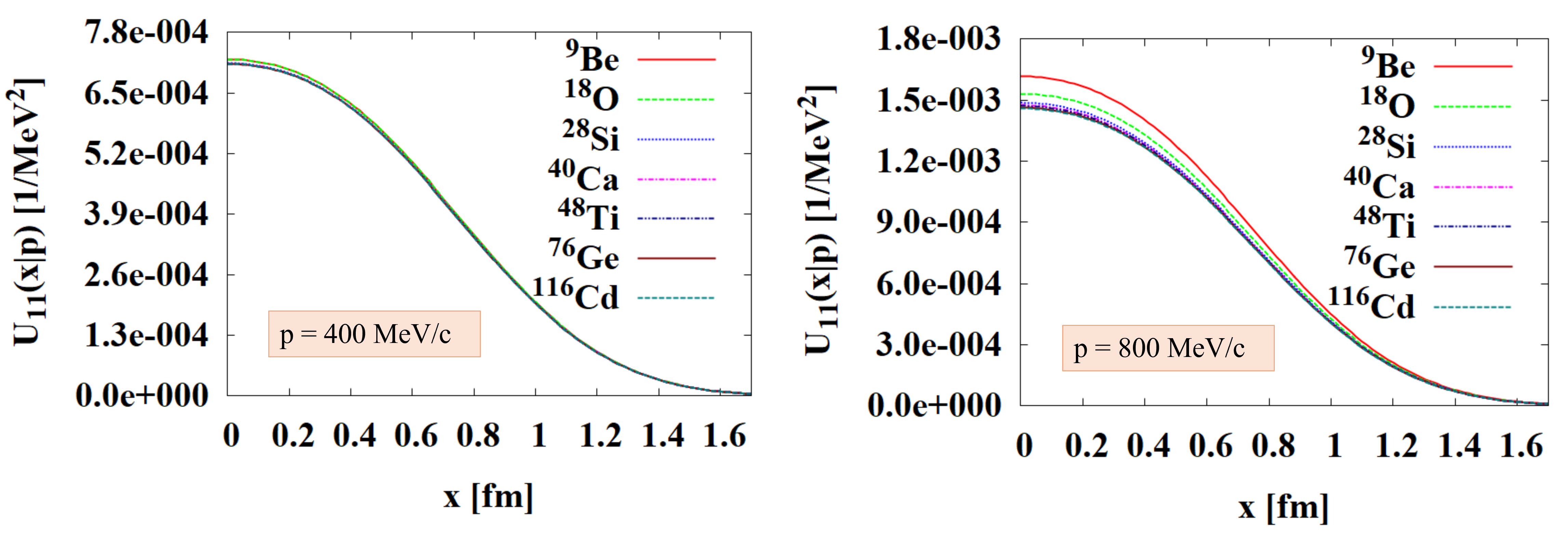}}
\caption{The mass dependencies of the $U_{11}$ component for a charged pion exchanged of momentum $|\vec{p}_1| = |\vec{p}_2| = 400$ MeV/c (left panel) and $|\vec{p}_1| = |\vec{p}_2| = 800$ MeV/c (right panel) is shown.}
\label{fig:universalityMDCE}
\end{figure}

Numerical calculations on pion potentials were performed at $T_{lab} = 270$ MeV using $^{18}$O projectiles on various target nuclei ranging from $^9$Be to $^{116}$Cd. These nuclei were selected as isotope candidates for $0\nu\beta\beta$ decay. A mild dependence of the potential components on the nuclear mass was found, slightly more pronounced at higher momenta of the charged pion exchanged in the MDCE reaction. A typical result of the mass dependence of the potentials is shown in Fig. \ref{fig:universalityMDCE}. These results provide numerical evidence for the universality of the MDCE process, as the pion potential appears to be nearly independent with respect to the nuclear masses involved in the reaction, particularly for medium and heavy nuclei.

\section{Conclusions}
The MDCE mechanism is a powerful probe of $0\nu\beta\beta$ decay dynamics, exhibiting a universal behaviour with respect to the nuclear masses involved. This universality provides clear evidence of the short-range character of the process, as the pion potentials appear almost blind to the size of the colliding nuclear systems.

\section*{Acknowledgments}
H. Lenske acknowledges financial support in part by INFN and DFG , grant Le439/7.

\bibliography{biblio}

  
  


  
%
%
%
%
%
%
%
%
%
%



\end{document}